\documentclass[twocolumn,showpacs,preprintnumbers,amsmath,amssymb,floatfix]{revtex4}

\usepackage{graphicx}
\usepackage{dcolumn}
\usepackage{bm}

\begin{document}

%

\let\a=\alpha      \let\b=\beta       \let\c=\chi        \let\d=\delta
\let\e=\varepsilon \let\f=\varphi     \let\g=\gamma      \let\h=\eta
\let\k=\kappa      \let\l=\lambda     \let\m=\mu
\let\o=\omega      \let\r=\varrho     \let\s=\sigma
\let\t=\tau        \let\th=\vartheta  \let\y=\upsilon    \let\x=\xi
\let\z=\zeta       \let\io=\iota      \let\vp=\varpi     \let\ro=\rho
\let\ph=\phi       \let\ep=\epsilon   \let\te=\theta
\let\n=\nu
\let\D=\Delta   \let\F=\Phi    \let\G=\Gamma  \let\L=\Lambda
\let\O=\Omega   \let\P=\Pi     \let\Ps=\Psi   \let\Si=\Sigma
\let\Th=\Theta  \let\X=\Xi     \let\Y=\Upsilon

%

%

\def\cA{{\cal A}}                \def\cB{{\cal B}}
\def\cC{{\cal C}}                \def\cD{{\cal D}}
\def\cE{{\cal E}}                \def\cF{{\cal F}}
\def\cG{{\cal G}}                \def\cH{{\cal H}}
\def\cI{{\cal I}}                \def\cJ{{\cal J}}
\def\cK{{\cal K}}                \def\cL{{\cal L}}
\def\cM{{\cal M}}                \def\cN{{\cal N}}
\def\cO{{\cal O}}                \def\cP{{\cal P}}
\def\cQ{{\cal Q}}                \def\cR{{\cal R}}
\def\cS{{\cal S}}                \def\cT{{\cal T}}
\def\cU{{\cal U}}                \def\cV{{\cal V}}
\def\cW{{\cal W}}                \def\cX{{\cal X}}
\def\cY{{\cal Y}}                \def\cZ{{\cal Z}}
%

\newcommand{\Ns}{N\hspace{-4.7mm}\not\hspace{2.7mm}}
\newcommand{\qs}{q\hspace{-3.7mm}\not\hspace{3.4mm}}
\newcommand{\ps}{p\hspace{-3.3mm}\not\hspace{1.2mm}}
\newcommand{\ks}{k\hspace{-3.3mm}\not\hspace{1.2mm}}
\newcommand{\des}{\partial\hspace{-4.mm}\not\hspace{2.5mm}}
\newcommand{\desco}{D\hspace{-4mm}\not\hspace{2mm}}



\title{Source of \boldmath{$CP$} Violation for Baryon Asymmetry of the Universe 
 }

\author{
Wei-Shu~Hou$^{1,2,3,4}$
 }
\affiliation{$^{1}$Department of Physics,
 $^{2}$Institute of Astrophysics,
 $^{3}$National Center for Theoretical Sciences,
 $^{4}$
      Leung Center for Cosmology and Particle Astrophysics,
       National Taiwan University,
       Taipei, Taiwan 10617
 }

\date{\today}

\begin{abstract}
\noindent 
The baryon asymmetry of the Universe requires $CP$ violation, but
the Standard Model falls short by a factor of $10^{-10}$ or more.
Starting from a clue at the $B$ factories, we point out that the
large Yukawa couplings of the sequential fourth generation
$t^\prime$ and $b^\prime$ quarks can provide enhancement by a
factor of over $10^{13}$, making the 2-3-4 generation quark sector
a viable source of $CP$ violation for the baryon asymmetry of the
Universe. With recent hints of large $\sin2\Phi_{B_s}$ in
$B_s^0$-$\bar B_s^0$ mixing from the Tevatron, the ultimate test
would be to discover the $t^\prime$ and $b^\prime$ quarks at the
Large Hadron Collider.
\end{abstract}

\pacs{
 11.30.Er,
 12.15.Ff,
 13.25.Hw,
 98.80.Bp}
\maketitle

The Big Bang created matter and antimatter equally, but today we
see only protons, neutrons and electrons in our Universe; the
baryon asymmetry of the Universe (BAU) seems 100\%. Baryogenesis,
the elimination of antimatter while leaving behind some matter, is
one of the most fundamental problems.

One prerequisite for BAU is~\cite{Sakharov} the violation of
charge-parity symmetry (CPV). Laboratory measurements of CPV so
far all confirm the Kobayashi-Maskawa (KM) source~\cite{KM} in the
Standard Model (SM)~\cite{PDG06}. But the KM mechanism is known to
fall short of what is needed for BAU by over 10 orders of
magnitude! While this definitely motivates continued search in the
laboratories, the $10^{-10}$ factor may seem insurmountable.
In this Letter we point out a possible enhancement without
changing the SM dynamics in any essential way: a sequential fourth
quark generation could bridge the $10^{-10}$ gap.

The baryon-to-photon ratio of our Universe became precisely known
with WMAP data on the Cosmic Microwave Background Radiation
(CMBR)~\cite{WMAP03},
\begin{eqnarray}
\frac{n_{\cal B}}{n_\gamma} = (5.1^{+0.3}_{-0.2}) \times 10^{-10}.
 \label{eq:WMAP}
\end{eqnarray}
For every baryon, there are $2\times 10^9$ photons in the
2.7$^\circ$K CMBR. We see no antibaryons, however, so
\begin{eqnarray}
 {\cal A}_{\rm BAU} \equiv
  \frac{n_{\cal B} - n_{\bar{\cal B}}}
       {n_{\cal B} + n_{\bar{\cal B}}} =
       100\%.
 \label{eq:BAU}
\end{eqnarray}
The mystery is not so much the elimination of antimatter, but why
a tiny fraction of matter, Eq.~(\ref{eq:WMAP}), remains.

In 1967, Sakharov wrote down~\cite{Sakharov} the three conditions
for generating BAU:
i) $\,$ baryon number violation;
ii) $\; C$ and $CP$ violation;
iii) deviation from thermal equilibrium.
%
Sakharov was influenced by the experimental discovery~\cite{CPK}
of $CP$ violation in the form of $K_L^0 \to \pi^+\pi^-$ decay
occurring at the $2\times 10^{-3}$ level. In the early 1970s,
before the first two fermion generations were even established,
Kobayashi and Maskawa (KM) noticed~\cite{KM} that, if one extends
to a third quark generation, the weak interaction could have a
{\it unique source} of CPV in the quark sector.
This mechanism became part of SM.

It is remarkable that the SM carries all the ingredients to
satisfy~\cite{KRS85,Cline06} the Sakharov conditions. Although
conserved at the classical level, baryon number is
violated~\cite{tHooft76} by the triangle anomaly. Remarkably, the
extreme suppression at zero temperature becomes
unsuppressed~\cite{KRS85} for temperature $T$ above electroweak
energies of order 100 GeV. For the second condition, the weak
interaction violates $C$ invariance, and the KM mechanism violates
$CP$ invariance. For the third condition, the electroweak phase
transition (EWPhT) could~\cite{KRS85} be strong enough to cause
deviation from equilibrium. We will return to this last point in
our discussions. Let us understand why the KM theory, which can
explain all $CP$ asymmetries in the laboratories, is $10^{-10}$
too small for BAU.

The gauge coupling $g$ of a $W^-$ boson to the $\bar u_i d_j$
quark pair is modulated by the quark mixing matrix element
$V_{ij}$. KM showed~\cite{KM} that the $2\times 2$ matrix is
orthogonal with no phase, but the $3\times 3$ matrix $V$ is
unitary, with a unique weak phase. Furthermore, in the three
generation KM theory (which we call SM3), if any two like charged
quarks are equal in mass, it effectively reduces to a two
generation theory with no phase. CPV involves all three
generations of quarks of both charges.

By invariance arguments, Jarlskog pointed out~\cite{Jarlskog} that
${\rm Im}\det\bigl[m_u m_u^\dag,\;m_d m_d^\dag\bigr]$ can be used
as the measure of CPV. The general CPV invariant in SM3 is
\begin{eqnarray}
J &=& (m_t^2 - m_u^2)(m_t^2 -  m_c^2)(m_c^2 - m_u^2) \nonumber \\
  & & (m_b^2 - m_d^2)(m_b^2 -  m_s^2)(m_s^2 - m_d^2)\, A,
 \label{eq:J3}
\end{eqnarray}
where $A$ is twice the area of any triangle formed from the
unitarity condition $V^\dag V = I$. The $db$ element of $VV^\dag$
is one such triangle~\cite{PDG06} probed at the $B$ factories,
\begin{eqnarray}
V_{u d}V^*_{u b} + V_{c d}V^*_{cb} + V_{t d}V^*_{t b} = 0,
 \label{eq:btod}
\end{eqnarray}
which is illustrated by the small triangle in Fig.~1. By the
uniqueness of the CPV phase in SM3~\cite{KM}, all possible
analogues to Eq.~(\ref{eq:btod}) give the same area. For example,
the rather squashed triangle $O$-$V_{u s}V^*_{u b}$-$S$ in Fig.~1,
corresponding to $V_{u s}V^*_{u b} + V_{c s}V^*_{cb} + V_{t
s}V^*_{t b} = 0$, is the same in area. From Eq.~(\ref{eq:J3}) we
see that $J$ vanishes if $A=0$, or if any pair of like charged
quarks are degenerate.

\begin{figure}[t!]
\vskip-0.5cm\hskip-0.2cm
\includegraphics[width=3.4in,height=2.5in,angle=0]{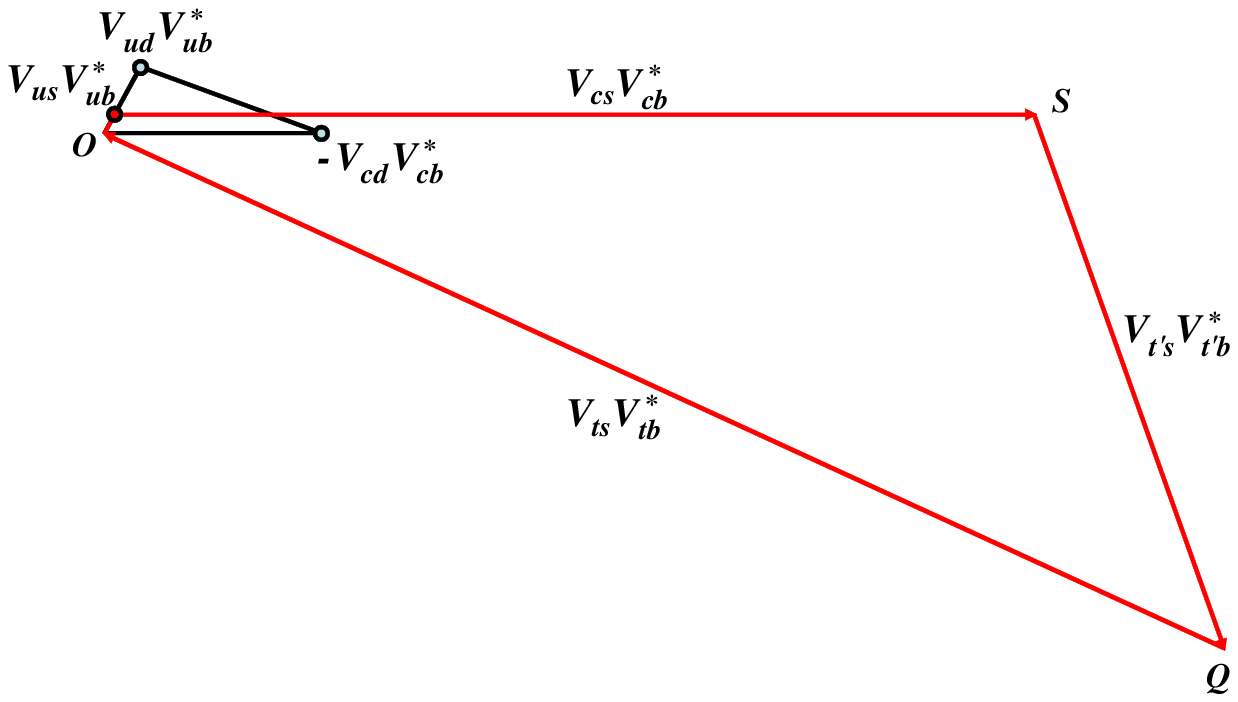}
\vskip-0.8cm
 \caption{
  Geometric representation of CPV in $b\to d$
  and $b\to s$ transitions.
  The small triangle represents our knowledge of
  the three generation unitarity relation, Eq.~(\ref{eq:btod}).
  The large quadrangle represents the four generation relation,
  Eq.~(\ref{eq:btos4}), drawn to scale with the $b\to d$ triangle,
  using~ $m_{t^\prime} = 300$ GeV to account for Eq.~(\ref{eq:DeltaAKpi}),
  as well as consistency with $Z^0$ and kaon data. See text for details.
  }
 \label{fig1}
\end{figure}

We can now see how CPV in SM3 falls far short of what is needed
for baryogenesis. As $J$ has 12 mass dimensions, normalizing by
the EWPhT temperature $T_{\rm EW} \sim 100$ GeV, together with $A
\simeq 3.1 \times 10^{-5}$~\cite{PDG06} one finds
\begin{eqnarray}
 J/T_{\rm EW}^{12}  \sim 10^{-20},
 \label{eq:J3T}
\end{eqnarray}
which falls short of Eq.~(\ref{eq:WMAP}) by $10^{-10}$ or
more~\cite{JoverT}.

SM3 can account for all CPV in the kaon and $B$ meson systems. In
face of Eq.~(\ref{eq:J3T}), many theories beyond SM3 that contain
large enough CPV for BAU have therefore been proposed~\cite{DK04}.
With recent advances in neutrino physics, the approach of
generating BAU through the lepton sector, i.e.
leptogenesis~\cite{FY86}, has gained in popularity.
We, however, aim to scrutinize the suppression in
Eq.~(\ref{eq:J3}) further. Our clue is the recent ``$\Delta{\cal
A}_{K\pi}$ problem" revealed by the B factories.

In 2004, direct CPV in the decay of $B^0$ vs $\bar B^0$ mesons was
established~\cite{belle04,babar04},
%
 ${\cal A}_{K^+\pi^-} \equiv {\cal A}_{B^0 \to K^+\pi^-} \cong -10\%$.
%
It is defined analogous to ${\cal A}_{\rm BAU}$, using the decay
rates of $\bar B^0 \to K^-\pi^+$ vs $B^0 \to K^+\pi^-$, and could
still arise from SM3. The Belle experiment recently
emphasized~\cite{belleDeltaA} a subtle, unexpected difference
between charged and neutral $B$ mesons, $\Delta{{\cal A}_{K\pi}}
\equiv {\cal A}_{K^+\pi^0} - {\cal A}_{K^+\pi^-} = +0.164 \pm
0.037$. The world average~\cite{HFAG} is now
\begin{eqnarray}
\Delta {\cal A}_{K\pi}
 & = & + 0.147 \pm 0.027,
 \label{eq:DeltaAKpi}
\end{eqnarray}
and well established. Although strong interaction effects cannot
be ruled out, ``it is equally possible that this is the first hint
of an entirely new mechanism for particle-antiparticle asymmetry",
through the so-called ``electroweak penguin"
process~\cite{Peskin}.
Though apparently a far cry, {\it does this offer new hope for
BAU?}

The difference $\Delta {\cal A}_{K\pi}$ is larger than the
measured strength of ${\cal A}_{K^+\pi^-}$. What new physics CPV
source could make such impact on the electroweak penguin amplitude
$P_{\rm EW}$? The SM3 contribution to $P_{\rm EW}$ is dominated by
the top quark, $P_{\rm EW}^{\rm SM3} \propto V_{ts}^*V_{tb}
f(m_t^2)$, which cannot affect $\Delta {\cal A}_{K\pi}$ because
$V_{ts}^*V_{tb}$ carries no weak phase in SM3. {Thus, new physics
is called for if $P_{\rm EW}$ is the culprit}. Note that the loop
function $f(m_t^2)$ grows as $m_t^2$ to first approximation,
rather than being suppressed by it: the top quark effect does not
decouple from $P_{\rm EW}$ for large $m_t$. Utilizing this unusual
{\it nondecoupling} behavior, we have advocated~\cite{HNS05} that
a natural possibility of generating $\Delta {\cal A}_{K\pi}$ is to
add a top-like $t^\prime$ quark: the new quark mixing element
product $V_{t^\prime s}^*V_{t^\prime b}$ carries a new CPV phase,
and the impact on $P_{\rm EW}$ grows with $m_{t^\prime}^2$! This
is still the KM theory, except one now has an extra, fourth quark
generation. We call this SM4.

\begin{figure}[t!]
\vspace{-1.6cm}
\includegraphics[height=2.5in]{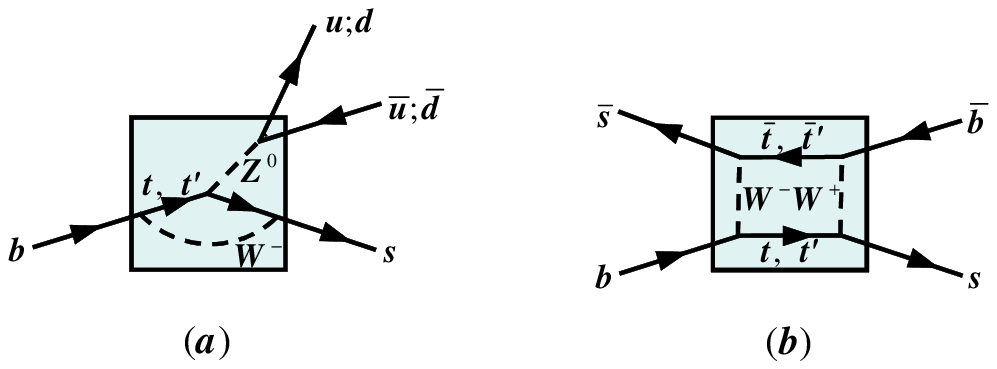}
\vskip-2.2cm
 \caption{
 The inner workings of
 (a) the electroweak penguin for $b\to sq\bar q$ transition, and
 (b) the box diagram for $\bar sb \leftrightarrow \bar bs$
     transitions.
 Dashed lines are used for $W^-$ and $Z^0$ bosons to indicate that the
 main effect comes from the longitudinal (or Goldstone boson) components.
 }
 \label{fig2}
\end{figure}

We illustrate the effect of $t$ and $t^\prime$ on $P_{\rm EW}$ in
Fig.~2(a). The $W$ boson loop around the $Z^0$ vertex converts the
$b$ quark into an $s$ quark, and the $Z^0$ boson turns into a
$\pi^0$. This is how $P_{\rm EW}$ enters $B^\mp\to K^\mp\pi^\pm$.
We have conducted analyses~\cite{HNS05,HLMN07} in SM4 using the
state-of-the-art perturbative QCD factorization approach, showing
that the $t^\prime$ contribution can account for
Eq.~(\ref{eq:DeltaAKpi}) through $P_{\rm EW}$. Detailed checks
were made~\cite{HNSprd05} on $Z\to b\bar b$ decay and kaon data,
finding that constraints~\cite{PDG06} on $V_{ij}$ are satisfied.
The outcome is illustrated in Fig.~1 for the nominal $m_{t^\prime}
= 300$ GeV (which is still consistent with Tevatron direct search
bounds~\cite{PDG06}).

The triangles of SM3 become quadrangles in SM4. For $b\to s$
transitions the relevant quadrangle is
\begin{eqnarray}
V_{u s}V^*_{u b} + V_{c s}V^*_{cb} + V_{t s}V^*_{t b}
  + V_{t^\prime s}V^*_{t^\prime b} &=& 0,
 \label{eq:btos4}
\end{eqnarray}
which is $O$-$V_{u s}V^*_{u b}$-$S$-$Q$ in Fig.~1. The strength of
$V_{t^\prime s}V^*_{t^\prime b}$, i.e. the vector
$\overrightarrow{SQ}$, which has a large angle with respect to
$V_{c s}V^*_{c b}$ (the vector from $V_{u s}V^*_{u b}$ to $S$
which is practically real), is a consequence of
Eq.~(\ref{eq:DeltaAKpi}). That is, the large phase and strength of
$V_{t^\prime s}V^*_{t^\prime b}$, together with the nondecoupling
of the $t^\prime$, generate the observed $\Delta {\cal A}_{K\pi}$.
The vector $\overrightarrow{QO}$, i.e. $V_{t s}V^*_{t b}$ in SM4,
is very different from $\overrightarrow{SO}$ in SM3.
Remarkably, the SM4 quadrangle of $V_{u d}V^*_{u b} + V_{c
d}V^*_{cb} + V_{t d}V^*_{t b} + V_{t^\prime d}V^*_{t^\prime b} =
0$ (not plotted), can barely be distinguished~\cite{HNSprd05} from
the small triangle of Eq.~(\ref{eq:btod}) for SM3. This explains
why there was no indication of deviation from SM3 prior to
Eq.~(\ref{eq:DeltaAKpi}).
Note that for $m_{t^\prime} >$ 300
GeV, the point $Q$ moves closer to $S$.

The quadrangle of Eq.~(\ref{eq:btos4}) actually mimics a triangle,
as $\vert V_{us}V_{ub}^*\vert$ is very small. The area, an
invariant measure of the strength of the $CP$ phase, is about 30
times the area $A$ of the SM3 triangle of Eq.~(\ref{eq:btod}). But
this pales against the factor of $10^{10}$ needed for generating
BAU, and one may despair.
We observe, however, that the small size of $J$ in
Eq.~(\ref{eq:J3}) is {\it due less to the $CP$ phase factor $A$,
but rather to the powers of light quark masses}, i.e.
$m_c^2m_b^4m_s^2/T_{\rm EW}^8 \sim 10^{-15}$. A similar point was
mentioned by Peskin~\cite{Peskin}. The masses of the $s$ and $c$
(and even $b$!) quarks are tiny compared to the electroweak
symmetry breaking scale of $v \simeq 246$ GeV. In terms of Yukawa
couplings $\lambda_i = \sqrt{2} m_i/v$, we have $\lambda_s \sim
0.0004$, $\lambda_c \sim 0.005$, and $\lambda_b \sim 0.017$.

Since it is the large Yukawa coupling, $\lambda_t \simeq 1$ for
$m_t \simeq 170$ GeV, that underlies the nondecoupling of the top
in $P_{\rm EW}$,
it may be the key to generating BAU as well. Shifting by one
generation, one replaces Eq.~(\ref{eq:J3}) by
\begin{eqnarray}
J_{(2,3,4)}^{sb}
 &\simeq& (m_{t^\prime}^2-m_c^2)(m_{t^\prime}^2-m_t^2)(m_t^2-m_c^2)
                                 \nonumber \\
 & &      (m_{b^\prime}^2-m_s^2)(m_{b^\prime}^2-m_b^2)(m_b^2-m_s^2)\,
                 A_{234}^{sb},   \nonumber \\
 &\sim&
    \frac{m_{t^\prime}^2}{m_c^2}
    \left(\frac{m_{t^\prime}^2}{m_t^2} - 1\right)
    \frac{m_{b^\prime}^4}{m_b^2m_s^2}\, \frac{A_{234}^{sb}}{A}\,J,
 \label{eq:J234}
\end{eqnarray}
in SM4, where $A_{234}^{sb} \simeq 10^{-3}$ is twice the area of
the large triangle by shrinking $\vert V_{us}V_{ub}^*\vert
\rightarrow 0$ in Fig.~1. The notation will be explained shortly.
Taking $m_t < m_{b^\prime} \lesssim m_{t^\prime} \sim$ 300 GeV,
{\it one gains 15 orders of magnitude}, with $10^{13}$ coming from
Yukawa couplings (i.e. quark masses)$\,$! Using $m_{b^\prime} \sim
m_{t^\prime} \sim 600$ GeV, one gains another factor of $10^2$;
the $\Delta {\cal A}_{K\pi}$ constraint on $\vert
V_{t's}V^*_{t'b}\vert m_{t^\prime}^2$ provides some control. In
the context of BAU, the approximations made to obtain
Eq.~(\ref{eq:J234}) are not so important compared to the striking
gain by many orders of magnitude.

Eq.~(\ref{eq:J234}) is not just a guess. CPV can in general be
written in terms of three-cycles, the trace of the cube of
commutators of quark masses~\cite{GKL86,Jarlskog87}. With four
generations, there are three~\cite{Jarlskog87} independent
sources, one related to $J_{(2,3,4)}^{sb}$ (we have modified the
more general notation of $J(2,3,4)$ of Jarlskog), another related
to $J$, which could have been written as $J_{(1,2,3)}^{db}$.
However, compared to $v\simeq 246$ GeV, $m_d \sim$ few MeV and
$m_s \sim 100$ MeV are close to massless. In the $d$-$s$
degeneracy limit, the three sources reduce to a single
one~\cite{Jarlskog87} (i.e. effectively 3 generation), which is
nothing but Eq.~(\ref{eq:J234}). Indeed, $J$ is suppressed by the
near degeneracy of $d$-$s$-$b$ as well as $u$-$c$, hence
vanishingly small compared to $J_{(2,3,4)}^{sb}$, which is
suppressed only by $m_b^2 - m_s^2$.

There should be two extra CPV phases~\cite{PDG06} compared to SM3.
Does $J_{(2,3,4)}^{sb}$ capture the dominant effect for BAU in
SM4? This is indeed the case. It is most easily seen by collapsing
the $b\to s$ quadrangle in Fig.~1 to a triangle, by shrinking
$V_{us}V^*_{ub}$ to point $O$. The change in area is small, and
the effective 2-4 generation world again has a unique $CP$ phase,
which is $A_{234}^{sb}$. Note that $J_{(2,3,4)}^{sb}$ by far
dominates over $J$, unless $A_{234}^{sb} \ll A$ by $10^{-13}$.
Thus, Eq.~(\ref{eq:J234}) gives the dominant effect of $CP$
violation relevant for BAU in SM4. {\it Baryogenesis is possible
with the dynamics that are already present in SM, and adding a
fourth quark generation realizes it.}

Some discussion is now in order.

First, our main result, the enhancement from replacing
Eq.~(\ref{eq:J3}) by Eq.~(\ref{eq:J234}), does not depend on
detailed values of $A_{234}^{sb}$, so long the latter does not
vanish. Thus, the starting point of Eq.~(\ref{eq:DeltaAKpi}), and
the subsequent discussion that lead to Eq.~(\ref{eq:J234}), are
just scaffolding that can be removed once the observation is made.

Second, why has the prominent enhancement by the fourth generation
gone unnoticed for so long? Since the 1990s, the fourth generation
is perceived as ruled out~\cite{PDG06} by electroweak precision
measurements (EWPM) and $N_\nu$ counting. But we now know the
neutrino sector is far richer than the naive SM, while the verdict
from EWPM has been contested recently~\cite{KPST07} (although
$|m_{t^\prime} - m_{b^\prime}|$ mass splitting is indeed
constrained). Our observation of a $10^{13}$ or more gain in CPV
argues in strong favor of SM4 over SM3, since Eq.~(\ref{eq:J3T})
shows that CPV in SM3 can never suffice for BAU.
We remark that the enhancement does not work for vector-like
exotic quarks, since their heaviness always lead to decoupling.

Third, most models with extra CPV for BAU tend to give too large
EDMs (electric dipole moments; see Ref.~\cite{Cline06} for some
discussion). But for SM4, the same mechanisms that keep EDMs small
in SM should be still at work.

Four, could a heavy fourth generation help bring about deviation
from equilibrium as well? It is remarkable that EWPhT could in
principle be strong enough in SM3 to satisfy Sakharov's third
condition, but the current Higgs mass bound rules out~\cite{DK04}
this possibility. It has therefore been popular~\cite{DK04} to
introduce extra heavy bosons that couple strongly to the Higgs
sector. There is recent speculation~\cite{CMQW05} regarding
whether fermions that couple strongly to the Higgs sector could
have similar effect. Though the top quark Yukawa coupling is not
large enough, a model of higgsino and wino with large Yukawa
couplings {\it could} strengthen the EWPhT into a first order one.
A very recent study, however, stresses~\cite{FK08} that something
similar is impossible for SM4 because of extra zero temperature
corrections. These authors then resort to~\cite{HOS05} $\tilde
t^\prime$ and $\tilde b^\prime$ squarks, which is again falling
back on the usual extra scalar boson approach.

In SUSY framework, one keeps all couplings perturbative, be it the
Higgs self-coupling $\lambda$, or Yukawa couplings $\lambda_i$.
However, we know that Nature does exploit strong, nonperturbative
effects, e.g. in QCD. For Yukawa couplings corresponding to
$m_{t^\prime}$, $m_{b^\prime}$ of order 600 GeV or higher,
unitarity violation sets in~\cite{CFH78}, and perturbation in
Yukawa couplings breaks down. This does not necessarily mean that
the theory ceases to exists, but rather, like in QCD, new bound
states appear to restore unitarity.
Such a picture has been advocated recently~\cite{Holdom06} for
even the breaking of electroweak symmetry itself, without the need
for a Higgs field (in other words, the Higgs becomes composite).
If such is the case in Nature, the issue of strength of EWPhT
should be revisited. This would depend on the actual high energy
theory, but it would be remarkable if a heavy fourth generation
could allow Standard Model dynamics to account for baryogenesis.
Such a picture, called electroweak
baryogenesis~\cite{KRS85,Cline06}, would be one of the most
beautiful outcomes of particle physics.

Five, our proposal offers exciting predictions that can be checked
in the next few years at the Tevatron and the Large Hadron
Collider (LHC). One prediction~\cite{HNS05,HNS07} is large
mixing-dependent $CP$ violation in the $B_s^0$-$\bar B_s^0$
system, in a measure called $\sin2\Phi_{B_s}$ that is akin to the
established SM3 effect in the $B_d^0$-$\bar B_d^0$
system~\cite{PDG06}. The $t$ and $t^\prime$ effect in the $\bar sb
\leftrightarrow \bar bs$ box diagram is illustrated in Fig.~2(b).
Like the $P_{\rm EW}$ amplitude of Fig.~2(a), these effects enjoy
nondecoupling. The SM3 prediction of $\sin2\Phi_{B_s} \simeq
-0.04$ is rather small~\cite{PDG06}, but our
prediction~\cite{HNS05,HNS07} of $\sin2\Phi_{B_s} \sim -0.5$ to
$-0.7$ for SM4 is rather striking. Recent reports from the
Tevatron~\cite{phisCDFtag,phisDzerotag} prefer the latter over SM.
By a combined fit to these and various Tevatron results, the UTfit
group~\cite{UTfit08} find a central value for $\Phi_{B_s} \sim
-0.67$, though it may be too early to claim evidence. The result
will certainly improve with more data. If $\sin2\Phi_{B_s}$ is
large, it can be quickly measured to good precision by the LHCb
experiment.

Note that a large and negative $\sin2\Phi_{B_s}$, though a
consequence of $\Delta {\cal A}_{K\pi} > 0$ in the four generation
model, is not a requirement for Eq.~(\ref{eq:J234}) to be realized
as the CPV source for BAU. Thus, the direct production of
$t^\prime$ and $b^\prime$ quarks are of even more interest. The
most recent Tevatron bound on $m_{t^\prime}$~\cite{tprime} is now
approaching the 300 GeV range. But whatever their masses, {\it the
$t^\prime$ and $b^\prime$ quarks can be readily discovered at the
LHC$\,$!} We will learn in just a few years time whether Nature
provides a fourth generation quark doublet with masses at several
hundred GeV, and with sufficient CPV for BAU.

Finally, we have not considered the associated 4th generation
heavy neutral and charged leptons, and the impact on neutrino
physics. If a fourth generation of quarks is discovered at the
LHC, the physics of the lepton sector would certainly be much
richer.

In summary, noting that CPV in the three generation Standard Model
is suppressed by the $s$, $c$ and $b$ quark masses hence too small
for baryogenesis, we point out that, adding a fourth generation of
quarks, one can gain a factor of $10^{13}$ or more. This could be
the source of $CP$ violation for generating the baryon asymmetry
of the Universe. Observation of large and negative
$\sin2\Phi_{B_s}$ could offer further support, and the direct
search for the fourth generation should be pursued vigorously at
the Tevatron and the LHC to test this scenario.

\vskip0.3cm
\noindent {\bf Acknowledgement}
 We thank F. Borzumati, T. Browder, P. Chang, L.-F. Li,
 E. Senaha, R. Sinha, A. Soni, C. Wagner, and R. Zwicky
 for discussions and advice.

\end{document}